\documentclass[10pt,amsfonts,amsmath,aps,pre]{revtex4}
\usepackage{graphicx}

\usepackage{array}

\newcommand*{\figany}[4]{
\begin{figure}[#1]
\begin{center}
     #2
   \caption{\label{#4}#3 $#4$
}
\end{center}
\end{figure}
}

\newcommand{\NEQU}[2]{
\begin{equation}
     #1
\label{#2}
\end{equation}
\centerline{\tm{\tiny Eq. (#2)}}
}
\newcommand{\EQU}[2]{
\begin{equation*}#1\end{equation*}
\centerline{\tm{\tiny Eq. (#2)}}
}

\newcommand{\WiePosit}{b}

\newcommand{\mb}[1]{\mathbf{#1}}

\newcommand{\tm}[1]{\textrm{#1}}

\newcommand{\ind}[1]{{\textrm{\tiny{#1}}}}
\newcommand{\eww}[1]{\la{#1}\ra}
\newcommand{\Int}[2]{\int\limits_{#1}^{#2}\d}

\newcommand{\bra}[1]{\left({#1}\right)}

\newcommand{\equ}[1]{Eq.~(\ref{#1})}

\newcommand{\figref}[1]{Fig.~\ref{#1}}

\newcommand{\rem}[1]{}

\renewcommand{\equ}[1]{Eq.~\ref{#1}}

\newcommand{\taua}{\tau_\alpha}

\newcommand{\pd}{\partial}

\renewcommand{\d}{{\rm d}}

\newcommand{\Eapp}{E_\ind{app}}

\newcommand{\eps}{\epsilon}
\newcommand{\epsmin}{\eps_\ind{min}}

\newcommand{\epsmb}{\eps_\ind{MB}}

\newcommand{\pos}{\mb{x}}

\newcommand{\la}{\left\langle}
\newcommand{\ra}{\right\rangle}
\renewcommand{\phi}{\varphi}

\newcommand{\bcfig}{\begin{figure}[H]\begin{center}}
\newcommand{\ecfig}{\end{center}\end{figure}}

\newcommand{\kB}{k_{\rm B}}

\newcommand{\NEQUARR}[2]{\begin{eqnarray} #1 \label{#2}\end{eqnarray}}

\renewcommand{\WiePosit}{!t}
\renewcommand{\paragraph}[1]{{\it #1}}
\renewcommand{\figref}[1]{figure~\ref{#1}}
\renewcommand{\equ}[1]{equation~\ref{#1}}

\renewcommand*{\figany}[4]{\begin{figure}[#1]\begin{center}     #2   \caption{\label{#4}#3
}\end{center}\end{figure}}
\renewcommand{\NEQU}[2]{\begin{equation}#1\label{#2}\end{equation}}
\renewcommand{\EQU}[2]{\begin{equation*}#1\end{equation*}}

\begin{document}

\title{Finite-Size Effects in a Supercooled Liquid.}

\author{Burkhard Doliwa* and Andreas Heuer\ddag
}

\address{*\ Max Planck Institute for Polymer Research, 55128 Mainz, Germany}
\address{\ddag\ Institute of Physical Chemistry, University of M\"unster, 48149 M\"unster, Germany}

\begin{abstract}
We study the influence of the system size on various static and dynamic properties of a
supercooled binary Lennard-Jones liquid via computer simulations.
In this way, we demonstrate that the treatment
of systems as small as $N=65$ particles yields relevant results for the understanding of
bulk properties.
Especially, we find that a system of $N=130$ particles behaves basically as two non-interacting
systems of half the size.
\end{abstract}

\maketitle

\section{Introduction.}

The theoretical understanding of the glass transition is still far from being complete.
During the last years, though, considerable progress has been made both from
the analytical and the numerical side (see, e.g., reference~\cite{Debenedetti:217}).
In this paper, we will dwell on some questions that arise within the framework
of the energy landscape approach~\cite{Goldstein:1969,Stillinger:333}.
The beauty of this viewpoint lies in the fact that the complicated nature
of many-particle effects in structurally disordered matter can be formulated
in a pictorial way by considering the topology of the high-dimensional landscape
of the total potential energy $V(x)$ (PEL).
(Here, $x=(\pos_1,...,\pos_N)$ denotes the set of positions of all $N$ particles in the system.)
Of special importance at low temperatures are the local minima of the PEL.
They are extremely numerous, so that a statistical treatment of their properties
is appropriate.
From the statistics of their energies and vibrational characteristics,
the whole thermodynamics can be derived at low-enough temperatures
and constant volume~\cite{Stillinger:222}.
Recently, generalizations to the constant-pressure situation and non-equilibrium
conditions have been numerically implemented~\cite{Mossa:391}.
The critical temperature $T_c$ of mode-coupling theory~\cite{Gotze:1992} serves
as a good indicator for the temperature range where the PEL standpoint
is appropriate: Below $T_c$ (the so-called landscape-dominated regime),
it is generally accepted that the temporal evolution
of a system happens through activated jumps among PEL minima.
Between $T_c$ and $2T_c$ (the landscape-influenced regime),
properties of minima are generally deemed to
be relevant for the thermodynamic description, whereas they
are expected to be irrelevant for dynamics there. This has been concluded from
the analysis of higher-order stationary points, which start to be populated
above $T_c$~\cite{Angelani:212,Broderix:228}.
In two recent publications, however, we have provided evidence that this
notion should be revised: From a detailed analysis of relaxation dynamics,
we have found that the picture of activated
hopping is correct even above $T_c$~\cite{doliwa:341,doliwa:392}.
In any event, above $2T_c$, the PEL description breaks down, due to the
fact that the system no longer occupies the well-behaved vicinity of minima.

In our recent publications,
we elucidated the implications of local PEL topology for
dynamics~\cite{Buchner:11,doliwa:341,doliwa:392}.
As conjectured by Stillinger~\cite{Stillinger:333}, we found that the PEL
is composed of groups of correlated minima, called metabasins~(MB).
Diffusional motion then turned out to consist of random jumps among metabasins,
where the jump times could be related to the depths of metabasins.
These conclusions were drawn from computational studies of fairly small model
systems of Lennard-Jones type (see reference~\cite{doliwa:392} for details about the systems).
The motivation for using systems as small as $N=65$ particles is two-fold:
Firstly, much longer time scales are accessible in the simulations,
and more sophisticated PEL analyses become possible~\cite{doliwa:392}.
Secondly, the global PEL viewpoint implies that the hypersurface of the potential energy
is the more complex the larger the system is. On the other hand, generally, relaxation
processes are spatially localized.
This implies that PEL complexity in large systems originates mainly from a
superposition of independently relaxing subsystems.
In contrast, the local relaxation dynamics itself is governed by
a non-trivial kind of PEL complexity which, however, is essentially
identical in all the subsystems.
Since we are interested in the physics behind local relaxation,
we concentrate on very small systems.
We thus avoid to consider many relaxing subsystems in parallel which would
average out the information about a single one.

In the present paper, we shall provide evidence that the results
obtained in reference~\cite{doliwa:341,doliwa:392}
for a small binary Lennard-Jones system of 65 particles (BMLJ65)
are also relevant for the bulk behavior.
Systems of $N=130,260$, and 1000 particles
will be investigated and compared to the BMLJ65.
Especially, we shall demonstrate that the BMLJ130 behaves essentially
like two non-interacting BMLJ65s.

We shall study static quantities in section~\ref{SECSTATIC}
and turn to dynamic observables in section~\ref{SECDYNAMIC}.
Further aspects of our results are discussed in section~\ref{SECCONCLUSIONS}.

\section{Static properties.}
\label{SECSTATIC}

\paragraph{Pair-correlation function.}
A first test for finite-size effects is to compare the distributions of interparticle distances $g_{\alpha\beta}(r)$
for different system sizes. Here, we restrict ourselves to the pair-correlation function of the
$A$-particles, $g_{AA}(r)$, see \figref{FIGFSE10}.
\figany{\WiePosit}{\includegraphics[width=10cm]{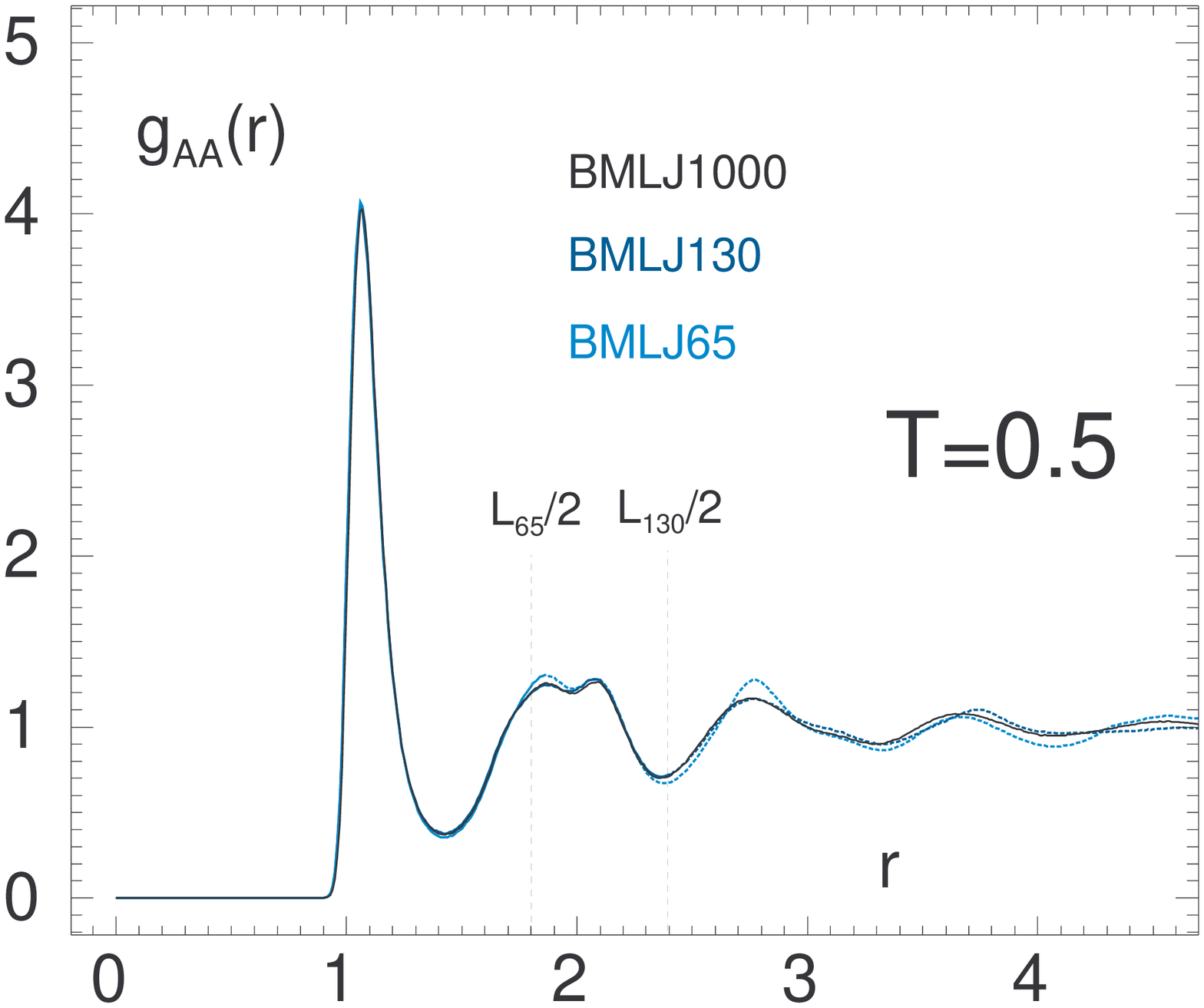}}{
Pair-correlation function, $g_{AA}(r)$, between $A$ particles,
for $N=1000,130,$ and 65.
Periodic images of the simulation box have been used to compute $g_{AA}(r)$
for distances larger than half the box width, $r>L_N/2$ ($N=130$ and $65$).
}{FIGFSE10}
Within a simulation box of width $L_N$, we may
only calculate $g_{\alpha\beta}(r)$ for $r<L_N/2$.
For larger values of $r$, periodic images of the simulation box must be used.
We find that $g_{AA}(r)$ of the BMLJ65 is identical to the one of the BMLJ1000 for
$r<L_{65}/2$. At larger distances, deviations from the bulk distribution can be seen.
This is plausible, since the simple duplication of the simulation box
can surely not reproduce all details of the long-ranged bulk correlations.
Nevertheless, the oscillations corresponding to higher-order neighbor shells in the BMLJ1000
are also present in the duplicated BMLJ65.
Similarly, the BMLJ130 matches the BMLJ1000 $g_{AA}(r)$ for $r<L_{130}/2$, whereas
deviations for larger $r$ already seem to be negligibly small.

\paragraph{Statistics of minima.}
It is an important question how the properties of the PEL are affected by changes
in system size. The most prominent characteristics of a PEL minimum are its energy $\eps$
and vibrational partition function $T^{(3N-3)/2}Y$. The latter can be calculated within harmonic
approximation,
\NEQU{Y=\prod_\nu\bra{\frac{2\pi}{\lambda_\nu}}^{1/2},}{EQUY}
where the $\lambda_\nu$ are the eigenvalues of the hessian matrix in the minimum.
Since the number of PEL minima is extremely large, a statistical treatment is needed.
As a starting point, we analyze the mean energy of minima at temperature $T$,
$\eww{\eps(T;N)}$, and their variance $\sigma^2(T;N)$.
For systems composed of independent subsystems,
$\eww{\eps(T;N)/N}$ and $\sigma^2(T;N)/N$ do not depend on system size.
\figany{\WiePosit}{\includegraphics[width=16cm]{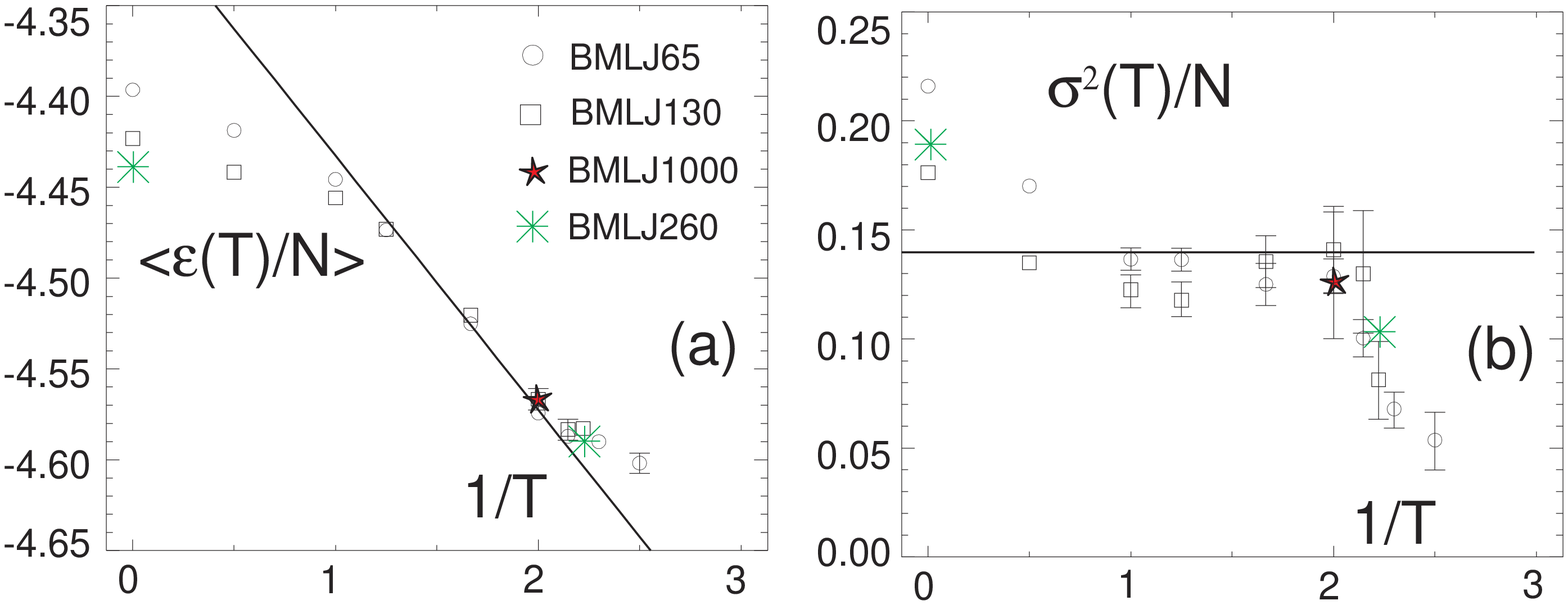}}{
(a) Mean minimum energy per particle vs. $1/T$.
Data of different system sizes are given.
(b) Variance of minimum energies, also on a per-particle base, again
vs. $1/T$.
In (a) and (b), the straight lines are the predictions from a gaussian density of minima.
At $1/T=0$, minimizations were performed from configurations with
random particle positions.
}{FIGFSE6}
In \figref{FIGFSE6}, these quantities are shown for $N=65$ and $N=130$, plus some
data points of $N=260$ and $N=1000$.
Concerning the mean energies, we find a good overall agreement of different system sizes.
The maximum difference is about $1\%$ between the BMLJ65 and the BMLJ260
at $T=\infty$.
In the landscape-influenced regime below $T=2T_c$, data for
different $N$ show a perfect match.
A similar conclusion can be drawn from \figref{FIGFSE6}(b), where we see $\sigma^2(T)/N$.
A systematically larger value is found for the BMLJ65 at high temperatures, as compared to
the BMLJ130. For $T\le 2T_c$, the difference is less than $20\%$, but
more precise statements are prohibited by
the statistical uncertainty of $\sigma^2(T)/N$ below $T=0.6$.
Thus, small but significant finite-size effects can be observed in this quantity.

We shortly comment on the deviations from the gaussian prediction at $T<0.5$,
as seen in \figref{FIGFSE6}(a),(b).
A possible explanation is that our simulations below $T=0.5$ were too
short to sample the PEL thoroughly at the lowest energies.
In view of the total simulation times for the BMLJ65
($T=0.4$: $50\taua$, $T=0.435$: $520\taua$, $T=0.45$: $850\taua$)
this is strong statement.
We currently study if significantly longer simulations could alter
the situation in \figref{FIGFSE6}, $T<0.5$.


\paragraph{Total number of minima.}
The overall statistics of PEL minima can be described to a large degree by
the number density of their energies, $G(\eps)$.
Formally, it can be written as $G(\eps)=\sum_i\delta(\eps-\eps_i)$, whereas
in practice, some coarse graining of energy is introduced.
In computer simulations, $G(\eps)$ was found to be approximately gaussian
for several model systems~\cite{Buchner:193,Sciortino:59,Starr:140,Emilia:302},
\NEQU{G(\eps)=\frac{1}{\sqrt{2\pi\sigma^2}}
\exp\bra{\alpha N-\frac{1}{2\sigma^2}(\eps-\eps_0)^2},
\quad\quad\eps>\epsmin.}{EQUGEPS}
The lower cutoff energy $\epsmin$ takes into account that the ideal gaussian cannot extend
to arbitrarily low energies.
A cutoff at the high-energetic tail of $G(\eps)$ is not needed,
due to the small Boltzmann weight of these minima.
For huge systems, which, to a good approximation, are composed of many independent
subsystems, the gaussianity of $G(\eps)$ is an immediate consequence of
the central limit theorem. However, as discussed in reference~\cite{Buchner:191},
a large degree of gaussianity must already be present in the subsystems that are considered
elementary.

The modification of the gaussian by the cutoff $\epsmin$ is normally small, so that one finds
$N_0(N)\equiv\int\d\eps G(\eps)\approx e^{\alpha N}$.
Thus, the so-called growth parameter $\alpha$ describes how the total number of minima, $N_0(N)$,
evolves with system size.
We will now calculate $N_0(65)$ and $N_0(130)$.
To this end, we apply a practical method which has recently been discussed
in the literature~\cite{Sciortino:280,Mossa:284,Sastry:198,Scala:71,Sastry:196}.
The main step is to compute the partition function $Z(T)$ or, equivalently, the
entropy via thermodynamic integration from a known reference state.
The knowledge of $Z(T)$ can be used to compute $G(\eps)$ as follows~\cite{Sastry:196}.
Using the harmonic approximation of basin vibrations,
the population of minimum~$i$
at temperature $T$ is
\NEQU{p_i=\frac{1}{Z(T)}T^{(3N-3)/2}Y_ie^{-\beta\eps_i},}{EQUPOP}
where $\beta=1/\kB T$.
By computing the expectation value $\sum_iY_i^{-1}e^{\beta\eps_i}\delta(\eps-\eps_i)p_i$,
we can then extract $G(\eps)$,
\NEQU{\ln G(\eps')=\ln\eww{\delta(\eps-\eps')Y^{-1}e^{\beta\eps}}+\ln Z(T)-\frac{3N-3}{2}\ln T.
}{EQUGEPSFROMTDI}

\figany{t}{
\includegraphics[width=14cm]{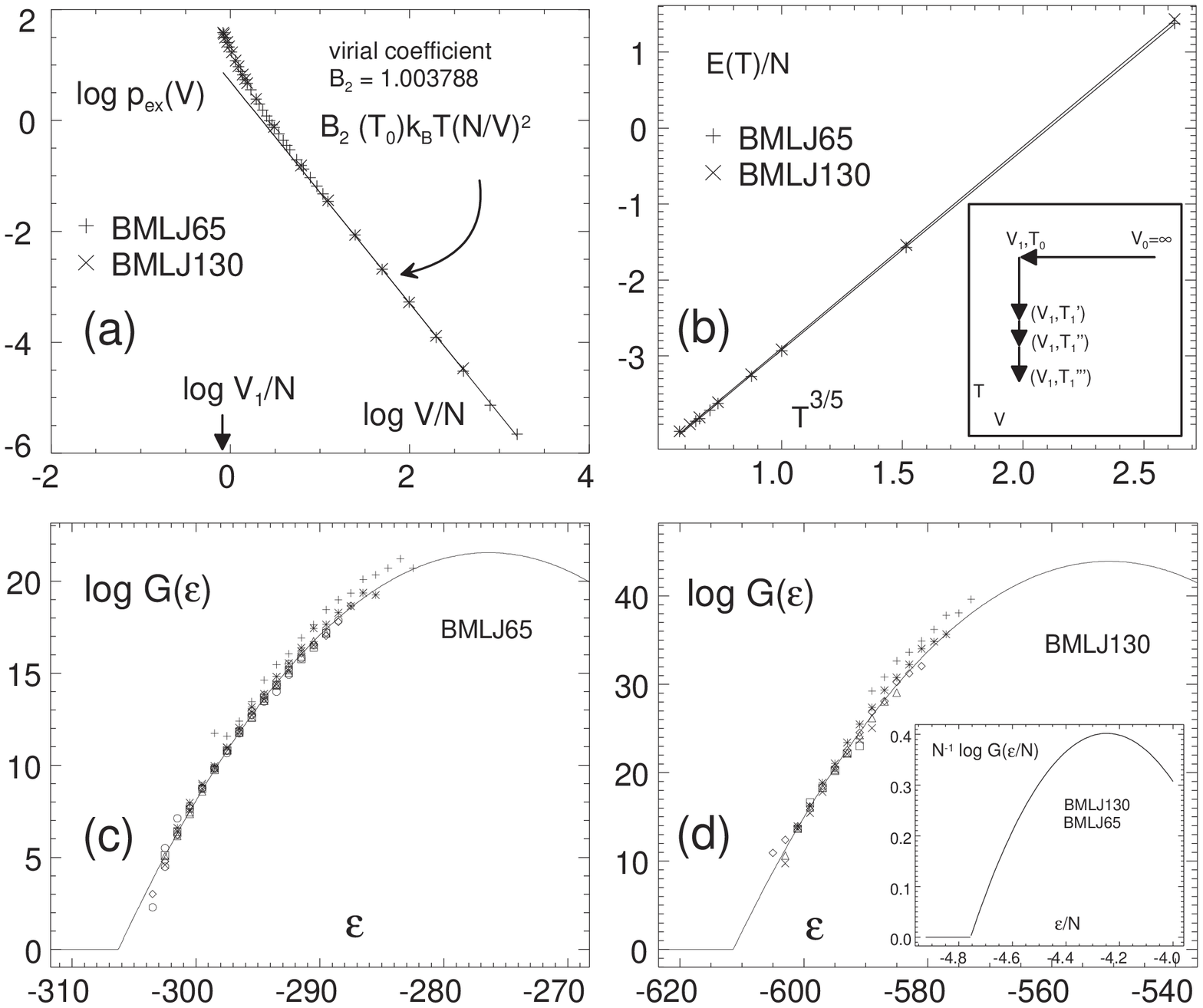}
}{
Determination of $G(\eps)$ via thermodynamic integration.
(a) Excess pressure $p_\ind{ex}=p-N\kB T/V$ over $V/N$ in a double-logarithmic plot.
The straight line corresponds to the first correction to ideal-gas behavior,
described by the second virial coefficient  $B_2(T)$.
(b) Temperature dependence of the mean potential energy $E(T)/N\equiv\eww{V(x)}/N$.
Lines are fits of the form $E(T)=a+bT^{3/5}$.
Note that the data of BMLJ65 and BMLJ130 practically coincide.
Inset: thermodynamic integration path in the $V-T$ plane ($T_0=5.0$, $N/V_1=1.2$).
(c) Number density of minimum energies, $G(\eps)$, computed via \equ{EQUGEPSFROMTDI}
from simulation runs of the BMLJ65 at $T=0.4,0.435,0.45,0.466,0.48,0.5,$ and $0.6$.
(d) Number density $G(\eps)$ of the BMLJ130, computed from
$T=0.4,0.435,0.45,0.5,$ and $0.6$.
In (c) and (d), data for $T\ge0.8$ ($+$) do not fall onto the master curve, indicating
the beginning breakdown of the harmonic approximation.
Inset: number density of minima of BMLJ65 and BMLJ130, on a per-particle base.
Curves coincide completely.
}{FIGFSE12}
In the above cited works, starting from a high-temperature ($T_0$), low-density
state ($N/V_0$), one compressed the system until the required volume $V_1$
of the supercooled liquid was reached.
Subsequently, one cooled along the isochore down to $T=T_1$.
The procedure is depicted in \figref{FIGFSE12}(b),inset.
The partition function at the state point $(V_1,T_1)$ follows with the help of the relations
\EQU{\bra{\frac{\pd\ln Z}{\pd V}}_T=\beta p(V,T),
\quad \tm{ and } \quad \bra{\frac{\pd\ln Z}{\pd\beta}}_V=-E(V,T),}{}
\NEQU{\ln Z(V_1,T_1)=\ln Z(V_0,T_0) +\beta_0\Int{V_0}{V_1}Vp(V,T_0)
- \Int{\beta_0}{\beta_1}\beta E(V_1,T).}{EQUTDI}
In the limit $V_0\to\infty$, we may replace $\ln Z(V_0,T_0)$ in \equ{EQUTDI} by the ideal gas term.
Note that we write $E(V,T)$ instead of $\eww{V}$ for potential energy here, in order
to avoid confusion with volume.

After measuring the pressure-volume dependence at $T_0=5$ (see \figref{FIGFSE12}(a)),
we evaluate the first integral of \equ{EQUTDI}.
This yields the partition function at $T_0=5$, $\rho_1=1.2$,
\NEQUARR{
\ln Z(V_1,T_0)&=-108.3\pm0.7 \quad\quad \tm{(BMLJ65)},\\
\ln Z(V_1,T_0)&=-218.5\pm2.5 \quad\quad \tm{(BMLJ130)}.
}{EQUTDIRES1}
Considering the second integral of \equ{EQUTDI},
we need the mean potential energy $E(T)$ along the isochore $V=V_1$ (see \figref{FIGFSE12}(b)).
As is done in the literature, we use the functional form $E(T)=a+bT^{3/5}$ to
parametrize our data. For the theoretical background of this form, see reference~\cite{Rosenfeld:70}.
From the two fit parameters
\NEQUARR{
a=-361.4\pm0.3,& \quad  b=171.5\pm0.3\quad\quad \tm{(BMLJ65)},\\
a=-718.8\pm0.8,& \quad  b=341.2\pm0.9\quad\quad \tm{(BMLJ130)},
}{EQUTDIRES2}
one can then calculate the second integral in \equ{EQUTDI}.

We now use these results to calculate $N_0$, via \equ{EQUGEPSFROMTDI}.
The expectation value in \equ{EQUGEPSFROMTDI} is
extracted from regular simulations.
Thus, as long as the harmonic approximation holds,
we are able to calculate the absolute value of $G(\eps)$.
For the BMLJ65 and BMLJ130 systems, $G(\eps)$ is shown in \figref{FIGFSE12}(c)
and~(d), respectively.
At temperatures $T\le0.6$, all points for $G(\eps)$  fall nicely onto a master curve,
whereas for $T\ge0.8$ the normalization does not work anymore, see reference~\cite{Sastry:196}.
We then fit gaussians to the data at $T\le0.6$, yielding the complete $G(\eps)$.
The number of minima can now be calculated from $N_0=\int\d\eps G(\eps)$,
\NEQUARR{
N_0&=10^{22.4\pm0.8}\quad \quad \tm{(BMLJ65)},\\
N_0&=10^{45.0\pm2.5}\quad \quad \tm{(BMLJ130)}.
}{EQUN0TDI}
Hence, within error bars, we find $N_0(130)=(N_0(65))^2$,
which is the trivial scaling behavior expected from
combinations of non-interacting subsystems.

We finally note that we reached the same conclusion after calculating the configurational
entropy $S_c(T;N)$ as described in reference~\cite{Sciortino:280} (data not shown here):
$S_c(T;N)/N$ turns out to be identical for $N=65$ and $N=130$.

\section{Dynamic properties.}
\label{SECDYNAMIC}
We now discuss the influence of system size on dynamics.
Here, more drastic effects than in static quantities are to be expected:
In fact, it is the most puzzling feature of the glass transition itself that a dramatic
slowdown of molecular motion cannot be traced back to changes in static quantities easily.

\paragraph{Diffusion coefficients.}
\figany{\WiePosit}{\includegraphics[width=10cm]{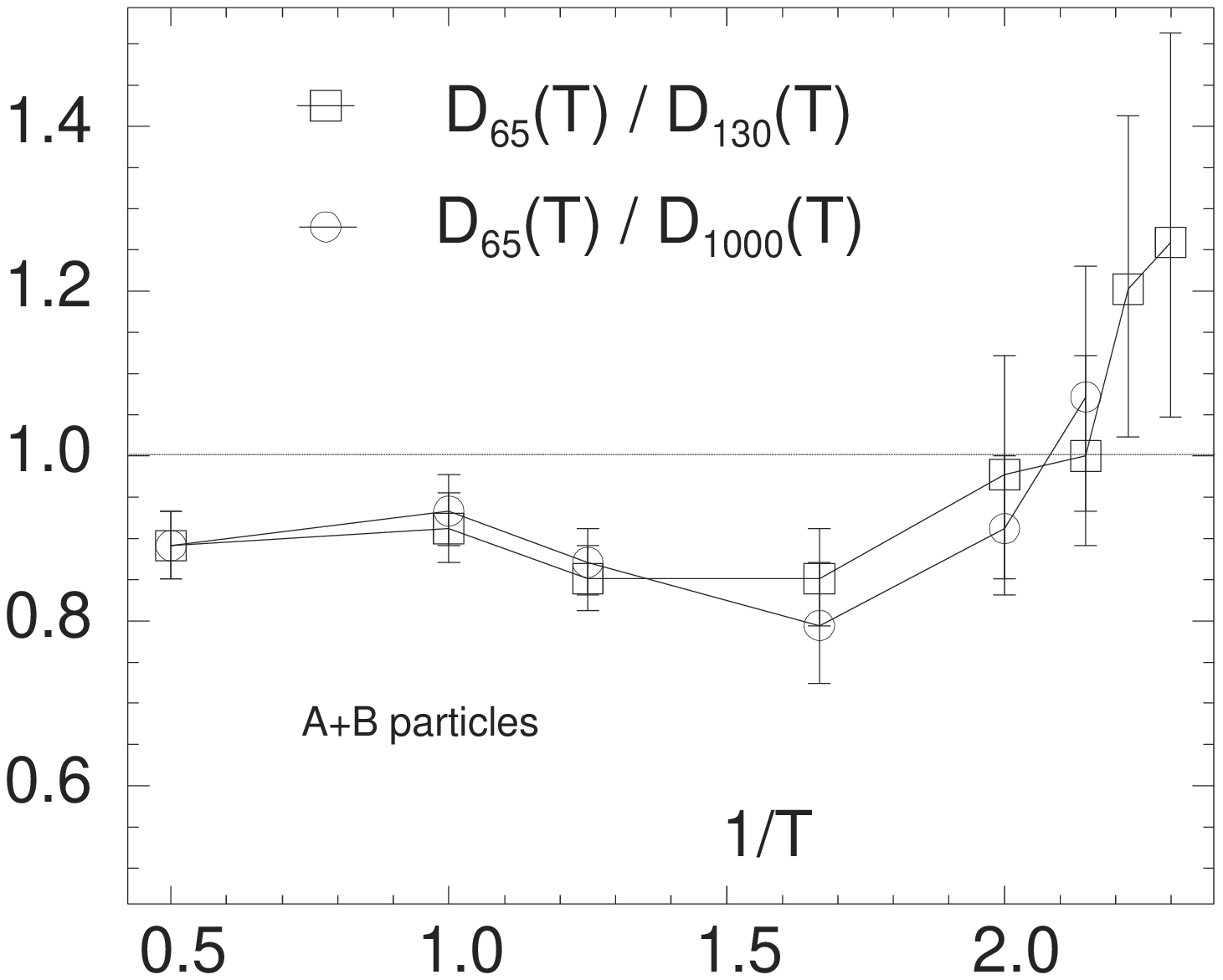}}{
Ratios of diffusion coefficients $D_N(T)$ for three system sizes versus $1/T$.
}{FIGFSE3}
We start with the long-time diffusion coefficient $D_N(T)$, defined by the Einstein relation.
One finds that the $D_N(T)$ for $N=65,130$, and $1000$ differ only very little.
In \figref{FIGFSE3}, we see $D_{65}/D_{1000}$ and $D_{65}/D_{130}$ as
functions of temperature.
The difference between $D_{65}$ and $D_{1000}$ -which we assume to be identical to
the bulk diffusion coefficient- is twenty percent or less above $T_c$.
Since data for $D_{1000}$ are not available below $T_c$, no such comparison is possible there.
As reflected by $D_{65}/D_{130}$, these differences are already present between
$N=65$ and $N=130$.
Below $T_c$, however, the BMLJ65 seems to become slightly faster than the BMLJ130.
However, since error bars are large for the two low-temperature data points,
it is hard to judge whether this is a systematic effect that further increases upon cooling.
In any event, in the temperature range studied,
the overall variation of $D_N(T)$ is more than three orders of magnitude.
Regarding the small deviation of the BMLJ65 relative to the BMLJ1000,
finite-size effects in the long-time diffusion should be jugded small.
As a comparison, major changes happen when going to $N=40$,
where we find $D_{40}(0.5)/D_{1000}(0.5)\approx0.1$.

\paragraph{Waiting-time distributions.}
As a more refined comparison of dynamics between different system sizes,
we consider the distributions of MB lifetimes (waiting times),
see references~\cite{doliwa:341,Reichman:398}.
A detailed description of MBs and their properties can be found in reference~\cite{doliwa:392}.
Here we briefly describe how MB lifetimes can be obtained from
a given simulation run.
Based on an equidistant time series of minima, we resolve the elementary transitions
between minima by further minimizations, accompanied by temporal
interval bisection.
From the series of configurations thus obtained, we determine the time intervals
of correlated back-and-forth jumps within groups of a few minima, which are
identified with the MB lifetimes~\cite{Buchner:11,doliwa:392}.
In this way, we are able to detect MB durations ranging from one MD step to
many millions of them.
At some arbitrary time of a simulation run,
the probability to be in a MB of length $\tau$
is $p(\tau)=\sum_i\tau_i\delta(\tau-\tau_i)/\sum_i\tau_i$,
where the $\tau_i$'s are the lifetimes found in the run.
(Since the MB residence times span more than six orders of magnitude,
our numerical computations will involve
the distributions $p(\log\tau)=p(\tau)\tau\ln10$ rather than $p(\tau)$.)
The temperature dependence of $p(\tau)$ will be suppressed for notational convenience.

\figany{\WiePosit}{\includegraphics[width=10cm]{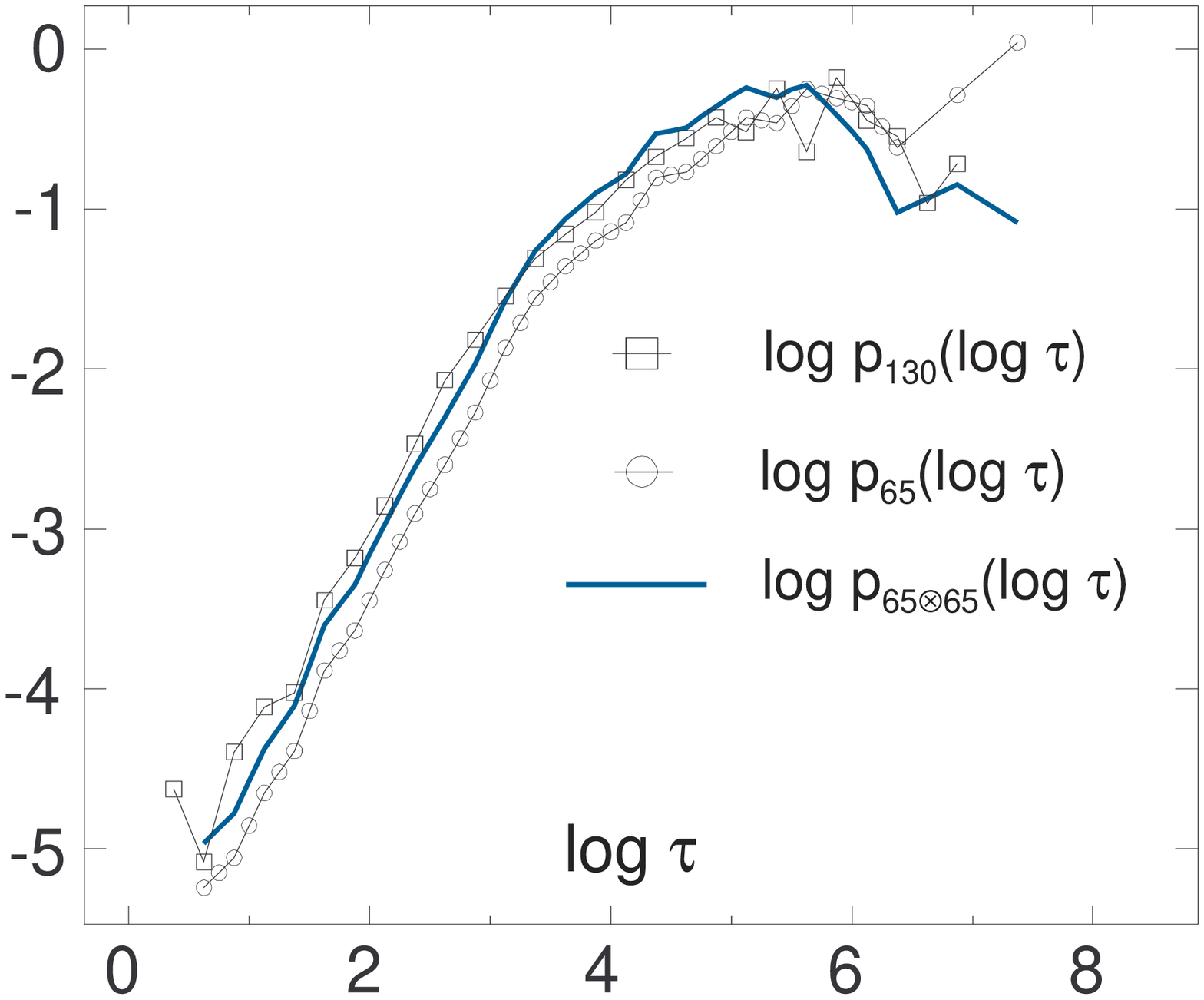}}{
Distributions of MB lifetimes, $p(\log\tau)$ at $T=0.5$.
As discussed in the text, the distribution $p_{130}(\log\tau)$
should be reproducable from $p_{65}(\log\tau)$
by a special kind of convolution, \equ{EQUCONVOLUTION}.
The corresponding function $p_{65\otimes65}(\log\tau)$
is given in the figure.
}{FIGFSE14}
Guided by the idea that a BMLJ130 system is basically a duplication of two independent
BMLJ65s, we may ask whether the distribution $p_{130}(\tau)$ of the larger system
can be reproduced by some sort of convolution of the distribution $p_{65}(\tau)$ of the smaller
one. (For a combined system, MB lifetimes are defined as the periods where
neither of the subsystems relaxes.)
Indeed, after a lengthy calculation one finds for the duplicated system,
\NEQU{p_{65\otimes65}(\tau)=-\frac{\d}{\d\tau}\Int{\tau}{\infty}{\tau'}
p_{65}(\tau')\Int{\tau}{\infty}{\tau''}p_{65}(\tau'')\bra{1-\frac{\tau^2}{\tau'\tau''}}.
}{EQUCONVOLUTION}
This expression can be simplified and, upon using $p(\log\tau)$, it reads
\NEQU{p_{65\otimes65}(\log\tau)=
2p_{65}(\log\tau)I(\tau)+2\tau\tilde I(\tau)\bra{\tilde I(\tau)\tau\ln10-p_{65}(\log\tau)},
}{EQUSIMPLER}
where
\EQU{I(\tau)=\Int{\tau}{\infty}{\tau'}p(\tau')\quad \tm{ and } \quad
\tilde I(\tau)=\Int{\tau}{\infty}{\tau'}p(\tau')/\tau'.
}{EQUSIMPLER2}
In \figref{FIGFSE11}, we show $p_{65}(\log\tau)$, together with $p_{130}(\log\tau)$
at the temperature $T=0.5$.
The distribution resulting from the duplication, $p_{65\otimes65}(\log\tau)$,
is also given in the figure. It agrees nicely with $p_{130}$.
Thus, on the refined level of waiting-time distributions, we find further evidence
that larger systems basically behave as consisting of non-interacting
BMLJ65-type building blocks.
Essentially, $p_{130}$ is shifted to the left with respect to $p_{65}$.
This is no wonder, because time intervals where both independent systems
are inert, are generally shorter than the waiting times of a single system.
For instance, the mean waiting times obey the relation
\EQU{\eww{\tau}_{65}=2\eww{\tau}_{65\otimes65},}{EQUMWT}
which can be shown with the help of \equ{EQUCONVOLUTION}.

\paragraph{Metabasin depths.}
\figany{\WiePosit}{\includegraphics[width=10cm]{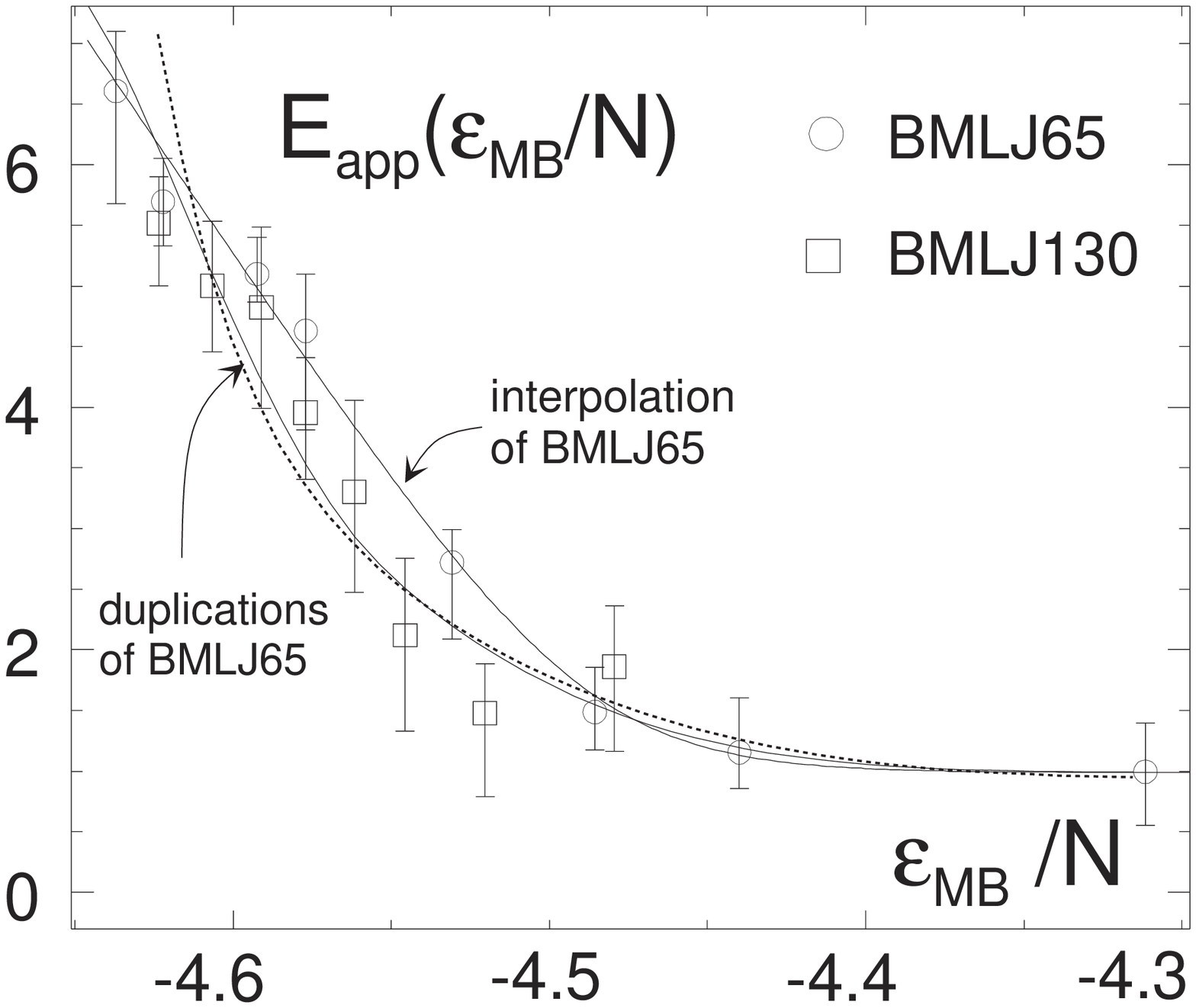}}{
Apparent activation energies $\Eapp(\epsmb/N)$ derived from mean lifetimes at fixed
metabasin energy $\epsmb$ (\equ{EQUTAUEPS}).
Data for $N=65$ and $N=130$ are shown versus $\epsmb/N$.
The interpolation of the $N=65$ data has been used to compute $\Eapp$
for the union of two non-interacting BMLJ65, as described in the text.
The result of this calculation is given in the figure (solid line).
Three further duplications yield $\Eapp$ for $N=1040$ (dashed line).
}{FIGFSE11}
As discussed above, metabasins have turned out as the relevant structures in the PEL
for describing the slowdown of molecular motion in supercooled liquids.
In a recent paper, we reported on how the average lifetimes $\eww{\tau|\epsmb;T}$
of MBs depend on their energies $\epsmb$~\cite{doliwa:392}.
At some fixed $\epsmb$, we found that $\eww{\tau|\epsmb;T}$ is Arrhenius-like
below $T\approx2T_c$, leading to the parametrization
\NEQU{\eww{\tau|\epsmb;T}=\tau_0(\epsmb)e^{\beta\Eapp(\epsmb)}.}{EQUTAUEPS}
The apparent activation energy $\Eapp(\epsmb)$ shows a strong dependence on $\epsmb$,
as soon as one drops below $\epsmb/N\approx-4.5$.
This can be seen in \figref{FIGFSE11} where $\Eapp(\epsmb)$ versus $\epsmb/N$
is depicted both for $N=65$ and $N=130$.
Naturally, it is tempting to relate $\Eapp(\epsmb)$ to PEL structure.
By a detailed investigation of escape paths and the barriers that
are overcome when escaping the MBs, we were able to reproduce $\Eapp(\eps)$
from the local topology of MBs.
Hence, we were able to prove that the effective depth $\Eapp$ of a MB, derived from
dynamics, directly corresponds to the real depth of that MB as given by the energy barriers
around it.
Moreover, escaping from deep MBs could be shown to involve actived jumps even above $T_c$.
What concerns the prefactor $\tau_0(\epsmb)$ of \equ{EQUTAUEPS}, no such
understanding in terms of PEL structure has been possible.
Fortunately, however, $\tau_0(\epsmb)$ turned out
to have a weak dependence on MB energy.
Thus, it may in good conscience be set to a constant~\cite{doliwa:392}.

Here we concentrate on the dependence of $\Eapp(\epsmb)$ on system size.
As can be seen from \figref{FIGFSE11},
the activation energies $\Eapp(\epsmb/N)$ of $N=65$ and $N=130$ are quite close.
However, the $N=130$ data for $\epsmb/N<-4.5$ show the tendency to fall
slightly below that of $N=65$.
We shall show that this trend can be understood again in terms of a simple duplication
of a BMLJ65.
\newcommand{\epsmbone}{\epsmb^\ind{(1)}}
\newcommand{\epsmbtwo}{\epsmb^\ind{(2)}}
Hence, we are interested in the combination of two independent BMLJ65 systems.
Consider a MB of energy $\epsmb=\epsmbone+\epsmbtwo$ in the combined system.
Then assume that its average lifetime can be expressed through
the lifetimes of both subsystems, i.e.
\NEQU{\frac{1}{\eww{\tau|\epsmbone,\epsmbtwo}_{65\otimes65}}=
\frac{1}{\eww{\tau|\epsmbone}_{65}}+\frac{1}{\eww{\tau|\epsmbtwo}_{65}}.
}{EQUEAPP}
Averaging over the population of $\epsmbone$ and $\epsmbtwo$ at constant $\epsmb$
then yields $\eww{\tau|\epsmb;T}$ for the combined system.
The mean lifetimes produced in this way are again Arrhenius-like below $2T_c$ (data not shown here).
Thus, data can again be fitted by a function of the form of \equ{EQUTAUEPS},
yielding $\Eapp(\epsmb)$ for the duplicated BMLJ65.
The result is shown in \figref{FIGFSE11} for $T=0.5$.
Again, the artificial BMLJ65 duplication reproduces the observations for the real
system of $N=130$ particles.

Finally, we note that further duplication of the BMLJ65 leads to an interesting result:
The activation energies from duplication nearly fall on top of each other for
$N\ge130$ and $\epsmb/N>-4.6$.
The example of sixteen non-interacting BMLJ65s ($N=1040$) is shown in \figref{FIGFSE11}.

\section{Discussion and Conclusions.}
\label{SECCONCLUSIONS}

For several static and dynamic observables, we have verified the factorization property
\EQU{\tm{ BMLJ130 }\approx\tm{ BMLJ65 }\otimes\tm{ BMLJ65 }.}{EQUFAC}
The BMLJ130, in turn, seems to be close to bulk behavior.
Some of the results presented here have already been
obtained in earlier work for a very similar Lennard-Jones type system~\cite{Buchner:193}.
Again, the conclusion can be drawn that binary Lennard-Jones systems of ca. 60 particles
are a very good compromise between the desired smallness needed for our PEL investigations
and the required absence of large finite-size related artifacts.
In this connection, the simulation results of Yamamoto and Kim on the standard binary
soft sphere mixture are of interest~\cite{Kim:32}. Comparing systems of $N=108$ and $1000$
particles above $T_c$,
the authors found the small system to be up to an order of magnitude slower than the large one.
These findings suggest a fundamental difference between the Lennard-Jones and soft-sphere
types of relaxation dynamics. Evidently, soft spheres exibit a larger length scale
of cooperative motion than do Lennard-Jones systems.
A possible explanation can be found in reference~\cite{Buchner:193}.

It is known from the study of cooperative length scales that they
increase with decreasing temperature~\cite{Donati:355,Bennemann:95,doliwa:spatial}.
Thus, at some lower
temperatures one might expect that 65 particles are no longer enough and
finite-size effects become visible. For a similar Lennard-Jones system
it has been shown that finite-size effects are reflected by
the fact that the bottom of the PEL is frequently probed~\cite{Buchner:193}.
In the present
case still longer simulations at lower temperatures have to be performed
to check whether also for $N=65$ the PEL bottom can be reached.  Then the
interesting question arises whether or not differences to the $N=130$
system become visible.

The results presented in this paper suggest that the essential physics of the supercooled BMLJ
is already contained in the system of $N=65$ particles.
For the temperatures under investigation here,
extended structures of collectively moving particles ('strings') have been reported
in large systems~\cite{Donati:98}.
It is important to know whether these structures are still present in the small systems considered here.
Work along this line is in progress.

\section*{Acknowledgements.}
We are pleased to thank H.W.~Spiess for valuable discussions.
This work has been supported by the DFG, Sonderforschungsbereich 262.

\section*{References.}

\bibliographystyle{apsrev}

\bibliography{fse}

\end{document}